\documentclass{article}
\usepackage{amsmath}
\usepackage[psamsfonts]{amssymb}
\usepackage{cmmib57}
\usepackage{diagrams}
\newcommand{\bPf}{\par\vspace*{-4pt}\indent{\sc Proof.}\enskip}
\newcommand{\ePf}{\medskip}
\def\QED{\hskip0.1em\hfill\null\ \null\nobreak\hfill\kern3pt\vbox{\hrule\hbox
   {\vrule\kern1pt\vbox{\kern1.7pt\hbox{$\scriptscriptstyle{QED}$}
    \kern0.2pt}\kern1pt\vrule}\hrule}}

\def\END{\hskip0.1em\hfill\null\ \null\nobreak\hfill\kern3pt\vbox{\hrule\hbox
   {\vrule\kern1pt\vbox{\kern1.7pt\hbox{$\,\,\,\vspace{5pt}$}
    \kern0.2pt}\kern1pt\vrule}\hrule}}
\def\con{{\offinterlineskip\lower 1truept\hbox{\kern2pt%
\vbox to7truept{\vfill\hbox to4truept{\hrulefill}}\vrule\kern3pt}}}
\newtheorem{theorem}{Theorem}
\newtheorem{lemma}{Lemma}
\newtheorem{corollary}{Corollary}
\newtheorem{proposition}{Proposition}
\newtheorem{remark}{Remark}
\newtheorem{definition}{Definition}
\newtheorem{example}{Example}
\newcommand{\bCd}{\bEq\begin{CD}}
\newcommand{\eCd}{\end{CD}\eEq}
\newcommand{\bcd}{\beq\begin{CD}}
\newcommand{\ecd}{\end{CD}\eeq}
\newcommand{\ben}{\begin{enumerate}}
\newcommand{\een}{\end{enumerate}}
\newcommand{\bEq}{\begin{eqnarray}}
\newcommand{\eEq}{\end{eqnarray}}
\newcommand{\beq}{\begin{eqnarray*}}
\newcommand{\eeq}{\end{eqnarray*}}
\newcommand{\bDf}{\begin{definition}\em}
\newcommand{\eDf}{\end{definition}}
\newcommand{\bLm}{\begin{lemma}}
\newcommand{\eLm}{\end{lemma}}
\newcommand{\bPr}{\begin{proposition}}
\newcommand{\ePr}{\end{proposition}}
\newcommand{\bTh}{\begin{theorem}}
\newcommand{\eTh}{\end{theorem}}
\newcommand{\bCr}{\begin{corollary}}
\newcommand{\eCr}{\end{corollary}}
\newcommand{\bRm}{\begin{remark}\em}
\newcommand{\eRm}{\end{remark}}
\newcommand{\bEx}{\begin{example}\em}
\newcommand{\eEx}{\end{example}}

\newcommand{\ie}{{\em i.e$.$} }
\newcommand{\eg}{{\em e.g$.$} }
\newcommand{\R}{I\!\!R}

\newcommand{\mto}{\mapsto}

\newcommand{\der}{\partial}

\DeclareMathOperator{\im}{im}
\DeclareMathOperator{\byd}{{\raisebox{.1ex}{:}{=}}}

\newcommand{\ucar}[1]{\underset{#1}{\times}}

\newcommand{\owed}[1]{\overset{#1}{\wedge}}

\newcommand{\balp}{\boldsymbol{\alp}}

\newcommand{\brho}{\boldsymbol{\rho}}
\newcommand{\bsig}{\boldsymbol{\sig}}

\newcommand{\cC}{\mathcal{C}}

\newcommand{\cE}{\mathcal{E}}
\newcommand{\cF}{\mathcal{F}}

\newcommand{\cL}{\mathcal{L}}

\newcommand{\bg}{\boldsymbol{g}}

\newcommand{\bQ}{\boldsymbol{Q}}

\newcommand{\bU}{\boldsymbol{U}}

\newcommand{\bX}{\boldsymbol{X}}
\newcommand{\bY}{\boldsymbol{Y}}

\newcommand{\sub}{\subset}

\newcommand{\wed}{\wedge}

\newcommand{\ten}{\!\otimes\!}

\newcommand{\alp}{\alpha}
\newcommand{\bet}{\beta}
\newcommand{\gam}{\gamma}
\newcommand{\del}{\delta}

\newcommand{\lam}{\lambda}
\newcommand{\sig}{\sigma}

\newcommand{\ome}{\omega}
\newcommand{\Gam}{\Gamma}
\newcommand{\Del}{\Delta}

\newcommand{\For}{{\Lambda}}
\newcommand{\Con}{{\mathcal{C}}}
\newcommand{\Hor}{{\mathcal{H}}}
\newcommand{\Var}{{\mathcal{V}}}
\newcommand{\bVar}{\bar{\mathcal{V}}}
\newcommand{\Thd}{{\Theta}}
\newcommand{\cprime}{\/{\mathsurround=0pt$'$}}

\let\abs=\envert

\newcommand{\hL}{\bar{\Lambda}}
\title{\textbf{The Hessian and Jacobi Morphisms \\
for Higher Order Calculus of Variations}}
\bigskip
\author{\large{Mauro Francaviglia, Marcella Palese}\thanks{%
Supported by GNFM of INDAM, MURST and Universities of
Torino and Lecce.}
\\{\footnotesize Department of Mathematics,
University of Torino}
\\{\footnotesize via C. Alberto 10, 10123 Torino, Italy}
\\{\footnotesize email: \texttt{francaviglia@dm.unito.it,
palese@dm.unito.it}}
\\
{\footnotesize and}\\
\large{Raffaele Vitolo$^*$}
\\{\footnotesize Department of Mathematics,
University of Lecce}
\\{\footnotesize via Arnesano, 73100 Lecce, Italy}
\\{\footnotesize email: \texttt{raffaele.vitolo@unile.it}}}
\date{}
\begin{document}

\maketitle

\begin{abstract}
We formulate higher order variations of a Lagrangian in the geometric
framework of jet prolongations of fibered manifolds. Our formalism applies to
Lagrangians which depend on an arbitrary number of independent and dependent
variables, together with higher order derivatives. In particular, 
we show that the second
variation is equal (up to horizontal differentials) to the vertical differential of the
Euler--Lagrange morphism which turns out to be self-adjoint along solutions of 
the Euler-Lagrange equations. These two
objects, respectively, generalize in an invariant way the Hessian morphism and
the Jacobi morphism (which is then self-adjoint along critical sections) of a given 
Lagrangian to the case of higher order Lagrangians.
Some examples of classical Lagrangians are provided to illustrate our method.

\noindent {\bf Key words}: fibered manifold, jet space, variational
sequence, second variation.

\noindent {\bf 2000 MSC}: 58A20, 58E30.
\end{abstract}
\newpage
\section{Introduction}\label{1}
An important aspect of mathematics that can be fit into differential
geometry is the calculus of variations. This research started with
several formulations of calculus of variations on jet spaces
(see, \eg, \cite{GoSt73,Kru73} and Appendix
2). Jet spaces are the natural framework for differential equations 
in differential geometry.
Good sources are \cite{Many99,KMS93,Kup80,MaMo83a,Olv93,Pal68,Sau89}.
It become later evident that the passage from Lagrangians to
Euler--Lagrange equations was nothing but a differential of a certain complex
\cite{Tak79,Tul77,Vin77,Vin78,Vin84}: this lead to variational sequences and
much more.

So far we found that an interesting aspect of calculus of variations
was not developed in all details from 
the point of view of geometric
formulations on jet spaces: namely, the second and higher variations of
Lagrangians. The Lagrangian characterization of the second variation of a
Lagrangian in the framework of jet bundles has been considered in 
\cite{GoSt73} and, more recently,
in \cite{CaFr97a,CaFr97b,CFT96,Cra00}. We stress that in
\cite{CaFr97a,CaFr97b,GoSt73} only first-order Lagrangians were considered,
while in \cite{CFT96} also a distinguished class of second-order Lagrangians
has been studied.  In particular, in \cite{CaFr97a,CaFr97b,CFT96} it was shown
how to recast (up to divergencies) the system formed by the Euler--Lagrange
equations together with the Jacobi equations for a given Lagrangian as the
Euler Lagrange equations for a \emph{deformed} Lagrangian.

In this paper we provide a geometrical characterization of the second
variation of a Lagrangian of arbitrary order in the general case of $n$
independent variables and $m$ unknown functions.
The second variation can be written in an infinite number of ways, by adding
arbitrary total divergencies. Some preliminary results concerning the Jacobi
morphism were obtained in \cite{FP01,FrPa02,Pa00}. In the present paper, we 
exhibit a distinguished
representative in the class of all such forms. 
Our representative has
remarkable intrinsic and coordinate interpretations. 
In particular, we prove that, within the framework of finite order variational
sequences, the Jacobi morphism turns out to be self-adjoint along critical sections, \ie, along solutions of the
Euler-Lagrange equations (see also \cite{PaWi03} for some important consequences of this fact). We show clearly the
connection between Hessian and Jacobi morphisms; furthermore, their representatives are ready for applications to
arbitrary order Lagrangians, as we show in the examples.

Notice that in literature only the first order case is usually treated: see \eg
\cite{BGG02,Cra00,GH96,Gri83}. Moreover, our formulation has the advantage to be
easily generalizable to higher variations. In \cite{Cra00} a Poincar\'e--Cartan
form is used in the first-order case and for one independent variable, but in
higher-order calculus of variations the Poincar\'e--Cartan form is no longer
unique, so that the approach would not lead to a unique formulation of
variations of higher order Lagrangians.  Last but not least, our 
approach can be further generalized to
higher variations of Euler--Lagrange type morphisms, Helmholtz morphisms and
to all forms of the variational sequence. This is relevant also in view 
of the role played by the second
variation in many geometric contexts. For example, the second 
variation of the Yang-Mills functional
has an algebraic
structure which leads in dimension $4$ to important geometric
consequences about stable Yang-Mills connections, such
as local minima of the functional \cite{AtBo83,Bou87}. As well as, it is 
relevant for
the study of the theory of stable and
unstable minimal submanifolds of a Riemannian manifold \cite{Ros97}.

The paper is organized as follows.  In section 2 we recall elements of calculus
of variations on jet spaces. We use the language of finite order variational
sequences, as developed by Krupka \cite{Kru90}.

In section 3 we introduce the notion of $i$-variation of a section as an
$i$--parameter `deformation'. The variations that we consider are of general
(nonlinear) form \cite{GH96}. This allows us to pass to infinitesimal
variations, \ie Lie derivatives, in a natural and straightforward way.  Then,
we introduce variation of forms as derivatives of their pull-back through the
variation of a section with respect to the parameters.

Such a notion of variation is applied in section 4. We concentrate ourselves on
the computation of the second variation of a Lagrangian.  It turns out that, on
critical sections (\ie on solutions of the Euler--Lagrange equations), the
second variation equals (up to total divergencies) the vertical differential of the Euler--Lagrange
morphism as well as its adjoint morphism. Thus we first of all generalize to higher order Lagrangians the well known fact
(in the first order case) that the second variation coincides with the Hessian morphism up to total divergences (see, \eg
\cite{GH96,GoSt73}); furthermore we characterize the
\emph{Jacobi
   morphism} of a given Lagrangian as the vertical differential of the Euler--Lagrange
morphism, which we show here to be self-adjoint along critical sections. We stress that, for a first order
Lagrangian, these two morphisms coincide with the standard Hessian and Jacobi
maps. Hence, our method generalizes maps related to the second variation to the
case of arbitrary order Lagrangians.

In the last section we consider a few well known and significant examples,
showing the role played in these cases by the geometric objects above.

Two short Appendices are included, in which we provide a synthetic 
version of the
well-known jet space formulation of variational problems, both for convenience
of the less experienced reader and for a better understanding of our motivation
and formalism.

\section{Variational sequences on jets of fibered manifolds}\label{sec:jets}

We recall in this section some basic definitions and results from the
theory of jet spaces. Complete treatments of this subject, with
different characters, can be found
in \cite{Many99,KMS93,Kup80,MaMo83a,Olv93,Pal68,Sau89}.
Our exposition follows more closely \cite{MaMo83a,Sau89}.

\medskip

Our framework is a \emph{fibered manifold} $\pi\colon \bY \to \bX$, with $\dim
\bX = n$ and $\dim \bY = n+m$. We recall that a fibered manifold is just a
surjective submersion $\pi$; in other words, fibers of $\pi$ need not to be
mutually diffeomorphic. A \emph{section} of $\pi$ is defined to be a map
$s\colon\bX\to\bY$ such that $\pi\circ s=id_{\bX}$.  We denote by
$V\bY\byd\ker T\pi\subset T\bY$ the \emph{vertical subbundle} of the tangent
bundle $T\bY$.
\subsection{Jet spaces}

For $r \geq 0$ we are concerned with the $r$--jet space $J_r\bY$.
This space is defined as the set of equivalence classes of sections of
$\pi$ having a contact of order at least $r$ at a given point. Equivalent
sections have the same $p$-th order differential at the given point,
$p\leq r$. We set $J_0\bY \equiv \bY$.

There are natural projections $\pi^r_s\colon J_r\bY \to J_s\bY$, $r
\geq s$, sending $r$-th equivalent sections into $s$-th equivalent
sections.  Moreover, there are obvious natural projections
$\pi^r\colon J_r\bY \to \bX$.  The spaces $J_r\bY$ are endowed with a
differentiable structure making $\pi^r_s$ fiber bundles.  Among these,
it can be proved that $\pi^r_{r-1}$ are affine fiberings ($r\geq 1$).

Charts on $\bY$ adapted to $\pi$ are denoted by $(x^\lam ,y^i)$.  Greek
indices $\lam ,\mu ,\dots$ run from $1$ to $n$ and they label base
coordinates, while Latin indices $i,j,\dots$ run from $1$ to $m$ and label
fiber coordinates, unless otherwise specified.  We denote by $(\der_\lam
,\der_i)$ and $(d^\lam, d^i)$ the local bases of vector fields and $1$--forms
on $\bY$ induced by an adapted chart, respectively.

We denote multi-indices of dimension $n$ by boldface Greek letters
such as $\bsig = (\sig_1, \dots, \sig_n)$, with $0 \leq \sig_\mu$,
$\mu=1,\ldots,n$;
we set $|\bsig| \byd \sig_{1} + \dots + \sig_{n}$ and $\bsig ! \byd
\sig_{1}! \dots \sig_{n}!$.
The charts induced on
$J_r\bY$ are denoted by $(x^\lam,y^i_{\bsig})$, with $0 \leq |\bsig|
\leq r$; in particular, we set $y^i_{\bf{0}} \equiv y^i$. The local
vector fields and forms of $J_r\bY$ induced by the above coordinates
are denoted by $(\der^{\bsig}_i)$ and $(d^i_{\bsig})$, respectively.

We recall that a section $s \colon\bX\to\bY$ can be prolonged to a section
$j_rs \colon \bX\to J_r\bY$. If we set $y^i\circ s = s^i$, then
we have the coordinate expression
\begin{displaymath}
(j_rs)^i_{\bsig}\byd y^i_{\bsig}\circ j_r s = \der_{\bsig}s^i\byd
\frac{\der^{\abs{\bsig}}s^i}{\der x^{\sigma_1}\cdots\der x^{\sigma_n}}.
\end{displaymath}
If the order of prolongation needed for $s$ in formulae is clear from context,
then we simply denote the prolongation of $s$ by $js$.

The jet spaces carry a natural structure, the \emph{Cartan
      (or contact) distribution} \cite{Many99}. It is the vector subbundle of
$TJ_r\bY$ generated by
vectors which are tangent to submanifolds of the form $j_r s(\bX)\subset
J_r\bY$. We present here a variant of this structure (see \eg \cite{MaMo83a}).

We consider the natural complementary fibered morphisms over the
affine fibering $J_{r+1}\bY \to J_{r}\bY$ induced by \emph{contact
      maps} on jet spaces
\beq
D : J_{r+1}\bY \ucar{\bX} T\bX \to
TJ_{r}\bY , \qquad \qquad \omega : J_{r+1}\bY \ucar{J_{r}\bY}
TJ_{r}\bY \to VJ_{r}\bY,
\eeq
with coordinate expressions, for $0 \leq
|\bsig| \leq r$, given by
\beq
D &= d^\lam\ten {D}_\lam = d^\lam\ten
(\der_\lam + y^j_{\bsig+\lam}\der_j^{\bsig}), \quad \omega &=
\omega^j_{\bsig}\ten\der_j^{\bsig} =
(d^j_{\bsig}-y^j_{{\bsig}+\lam}d^\lam) \ten\der_j^{\bsig}.
\eeq
Here,
the map $D$ is the inclusion of $T\bX$ into $TJ_{r}\bY$ through the
differential $Tj_{r}s$ of any prolonged section $j_rs$, while
$\omega\byd id_{TJ_{r}\bY}-D$. The vector field $D_\lambda$ is said
to be the \emph{total (or formal) derivative}; the forms
$\omega^j_{\bsig}$ are said to be \emph{contact (or Cartan) forms}.
Contact forms annihilate all vectors generated by $D_\lambda$. These
are tangent to submanifolds of the form $j_r s(\bX)\subset J_r\bY$.
We will use iterated total derivatives. Namely, if $f: J_{r}\bY \to
\R$ is a function, then we set $D_{\bsig,\lam}f$ $\byd D_{\lam} D_{\bsig}f$.

We have the following natural fibered splitting
\bEq
\label{jet connection}
J_{r+1}\bY\ucar{J_{r}\bY}T^*J_{r}\bY =\left(
      J_{r+1}\bY\ucar{\bX}T^*\bX\right) \oplus\Con^{*}_{r}[\bY],
\eEq
where
$\Con^{*}_{r}[\bY] \byd \im \omega^*$ is a subbundle of
      $J_{r+1}\bY\times_{J_{r}\bY}T^*J_{r}\bY$, and is naturally isomorphic to
$J_{r+1}\bY \times_{J_{r}\bY} V^{*}J_{r}\bY$ (see \cite{MaMo83a,Sau89}).

The above splitting induces splittings in the spaces of forms
\cite{Vit98}; here and in the sequel we implicitly use
identifications between spaces of forms and spaces of bundle morphisms
which are standard in the calculus of variations (see, \eg
\cite{Kol83,KMS93,Kru73}). Namely, let
$\For^{p}_r$ be the sheaf of
$p$--forms on
$J_r\bY$. We introduce the sheaves of \emph{horizontal forms}
$\Hor^{p}_{r+1,r}$, \ie, of fibered morphisms over
$\pi^{r+1}_{r}$ and $\pi^{r}$ of the type $\alp\colon J_{r+1}\bY \to
\wed^{p}T^*J_r\bY$ and $\bet\colon J_r\bY \to \wed^{p}T^*\bX$,
respectively. Finally, for $s\leq r$ we introduce the sheaves of
\emph{contact forms} $\Con^{p}_{r,s}$, \ie, of fibered
morphisms over $\pi^{r}_{s}$ of the type $\alp\colon J_{r}\bY\to
\wed^p\Con^{*}_{s}[\bY])$.

The splitting \eqref{jet connection} yields naturally the sheaf
splitting
\bEq\label{eq:wed_split} \Hor^{p}_{r+1,r}=
\bigoplus_{t=0}^p\Con^{p-t}_{r+1,r} \wed\Hor^{t}_{r+1}.
\eEq
Pull-back yields the inclusion $\For^{p}_r\subset\Hor^{p}_{(r+1,r)}$.
The effect of \eqref{eq:wed_split} on $\For^{p}_r$ is the following
\bEq\label{eq:wsplit_rest}
\For^{p}_r \sub\bigoplus_{t=0}^{p} \Con^{p-t}_r\wed\hL^t_r.
\eEq
Here, $\hL^t_r\byd h(\For^{p}_r)$ for $0 < p\leq n$ and $h$ is defined
to be the restriction to $\For^{p}_{r}$ of the projection of the above
splitting onto the non--trivial summand with the highest value of $t$.
Moreover, $\Con^{p-t}_r$ is the space of contact forms with values in
$\Con^{*}_{r}[\bY]$ (which is a bundle over $J_{r+1}\bY$) and
coefficients on $J_r\bY$. We define also the map $v \byd id - h$.

In other words, \emph{if $\alpha$ is a form on $J_r\bY$, then its pull-back
$(\pi^{r+1}_r)^{*}\alpha$ can be split into a part containing top degree
horizontal forms and a part containing more contact factors} (see,
\eg, \cite{Kru90,Vit98}).

In coordinates this is achieved by means of the substitutions
\beq
d^\lam\to d^\lam,\qquad d^i_{\bsig}\to
\omega^i_{\bsig}+y^i_{\bsig+\lam}\, d^\lam.
\eeq
which allow to express $\alpha$ in the basis $(d^\lam,
\omega^i_{\bsig})$ at the cost of raising the order of jet.

\medskip

The above splitting induces also a decomposition of the exterior differential
on $\bY$, $(\pi^{r+1}_r)^*\circ d = d_H + d_V$, where $d_H$ and $d_V$ are
called the \emph{horizontal} and \emph{vertical differential}, respectively.
The action of $d_H$ and $d_V$ on functions and $1$--forms on $J_r\bY$ uniquely
characterizes $d_H$ and $d_V$ (see, \eg, \cite{Sau89} for more
details). In particular, we have the coordinate expressions
\begin{gather*}
       d_Hf = D_\lam f\,d^\lam =
       (\der_\lam f + y^i_{\bsig+\lam}\der_i^{\bsig}f)d^\lam,\qquad
        d_Vf = \der_i^{\bsig} f \omega^i_{\bsig},
\\
       d_Hd^\lam = 0 ,\qquad d_Hd^i_{\bsig} = -d^i_{\bsig+\lam}\wed d^\lam ,
       \qquad d_H\omega^i_{\bsig} = -\omega^i_{\bsig+\lam}\wed d^\lam ,
\\
       d_Vd^\lam = 0 ,\qquad d_Vd^i_{\bsig}=d^i_{\bsig+\lam}\wed d^\lam ,
       \qquad d_V\omega^i_{\bsig} = 0 .
\end{gather*}

Any fibered isomorphism $F\colon \bY\to \bY$ over $id_{\bX}$
admits a prolongation to a fibered isomorphism $J_rF\colon J_r\bY\to
J_r\bY$ such that $J_rF\circ j_rs=j_r(F\circ s)$.

A vector field $\xi$ on $\bY$ is said to be \emph{vertical} if it has values
into $V\bY$. A vertical vector field can be conveniently prolonged to a
vertical vector field $j_{r}\xi\colon J_r\bY\to VJ_r\bY$. The vector field
$j_r\xi$ is characterized by the fact that its flow is the natural prolongation
of the flow of $\xi$. In coordinates, if $\xi=\xi^i\der_i$ we have $j_{r}\xi =
D_{\bsig}\xi^i\,\der_i^{\bsig}$, $0 \leq |\bsig | \leq r$.  Again, if the order
of prolongation needed in formulae for $\xi$ is clear from context, then we
simply denote the prolongation of $\xi$ by $j\xi$.

\medskip

Let $\alpha\in \Con^1_r\otimes\hL^n_r$. Then we can
interpret $\alpha$ as the differential operator
\[
\nabla_{\alpha}\colon \varkappa_0\to \hL^n_r,\quad \xi \mapsto j\xi\con\alpha,
\]
where $\varkappa_0$ is the space of vertical vector fields $\xi\colon \bY\to
V\bY$ and $\rfloor$ denotes the inner product. This is an operator in total
derivatives and its coordinate expression is
$\nabla_{\alpha}(\xi^i\der_i)=D_{\bsig}\xi^i\alpha_i^{\bsig}\, v_{\bX}$, where
$v_{\bX}\byd dx^1\wedge\ldots\wedge dx^n$ is the local volume form induced by a
chart $(x^\lambda)$ on $\bX$.  We can form the \emph{adjoint operator}
$\nabla_\alpha^*\colon (\hL^n_r)^*\otimes\hL^n_r=\cF_{r}\to
\varkappa_0^*\otimes\hL^n_r$.  It admits an intrinsic definition (see
\cite{Many99} for details). Its coordinate expression is
$\nabla_\alpha^*(f)=(-1)^{|\bsig|}D_{\bsig}(\alpha_i^{\bsig}f)\,
\omega^i\otimes v_{\bX}$.  This notion can be extended to a form
$\alpha\in\Con^1_r\otimes\Con^1_0\otimes\hL^n_r$: in this case we have
$\nabla_\alpha$, $\nabla_\alpha^*\colon\varkappa_0\to\Con^1_0\otimes\hL^n_r$,
with coordinate expression
\begin{equation}\label{adjoint}
\nabla_\alpha^*(\xi^i\der_i)=
(-1)^{|\bsig|}D_{\bsig}(\alpha_{ij}^{\bsig}\xi^j)\,\omega^i\otimes v_{\bX}.
\end{equation}
\subsection{Variational sequence}
We recall now shortly the theory of variational sequences on finite order jet
spaces, as it was developed by Krupka in \cite{Kru90}.

Denote by
$(d\ker h)^s$ the sheaf generated by the presheaf $d\ker h$.  Set
$\Thd^{*}_{r}$ $\byd$ $\ker h$ $+$ $(d\ker h)^s$. Then the restriction of
exterior differential yields a sheaf sequence $0\to\Thd^*$ which is an
exact subsequence of the de Rham sequence of differential forms on $J_{r}\bY$.
\cite{Kru90}. Such a subsequence is made by forms that do not produce
contribution in action-like functionals \cite{GoSt73,Kru73,Olv93,Pal68}.

\bDf
The quotient sequence
\beq \diagramstyle[size=0.7em]
\begin{diagram}
      0 & \rTo & \R_{\bY} & \rTo & \For^{0}_r & \rTo^{\cE_{0}} &
      \For^{1}_r/\Thd^{1}_r & \rTo^{\cE_{1}} & \dots & \rTo^{\cE_{N-1}} & \For^{N}_r/\Thd^{N}_r & \rTo^{\cE_{N}} &
      \For^{N+1}_r & \rTo^{\cE_{N+1}} & \dots & \rTo^{d} & 0
\end{diagram}
\eeq
of the de Rham sequence with respect to the contact sequence is called the
$r$--th order \emph{variational sequence} associated with
the fibered manifold $\bY\to\bX$. Here the integer $N$ depends
on the dimension of the Cartan distribution on $J_{r}\bY$ \cite{Kru90}.\END
\eDf

The variational sequence is locally exact and, due to the `abstract'
de Rham Theorem, it has the same cohomology as the
de Rham sequence (see \cite{Kru90}).

Pull-back of forms through $\pi^{r+1}_r$ yields a natural inclusion of the
$r$-th order variational sequence into the $(r+1)$-th order variational
sequence. This allows us to represent any equivalence class
$[\alpha]\in\For^k_r/\Thd^k_r$ with a single morphism of bundles over jet
spaces, at the cost of raising the order $r$. More precisely, the quotient
sheaves $\For^k_r/\Thd^k_r$ in the variational sequence are represented as
sheaves of bundle morphisms $\Var^k_r$ (defined on jets of order $s>r$) via the
following \emph{intrinsic} isomorphisms \cite{Vit98,Vit01}
\begin{subequations}
\begin{align}
      &I_{k}\colon \For^{k}_{r}/\Thd^{k}_{r} \to \Var^{k}_{r}\colon [\alp]
      \mto h(\alp), \qquad k\leq n,
      \\
      &I_{n+1}\colon \For^{n+1}_{r}/\Thd^{n+1}_{r}\to \Var^{n+1}_{r}\colon
      [\bet ]\mto E_{h(\bet )},
      \\
      &I_{n+2}\colon \For^{n+2}_{r}/\Thd^{n+2}_{r}\to \Var^{n+2}_{r}\colon
      [\gam ]\mto H_{h(\gam )}.
\end{align}
\end{subequations}

Let us describe the above morphisms $h(\alpha)$, $E_{h(\bet )}$,
$H_{h(\gam )}$ and spaces $\Var^h_r$, $h\leq n+2$.
\begin{enumerate}
\item $h(\alpha)$ is just the horizontalization of $\alp$; in the
    case $k=n$ the morphism $h(\alpha)$ can be interpreted as a
    Lagrangian density
    \cite{AnDu80,GoSt73,Kol83,KMS93,Kru73,Kup80,Pal68,Tul77}.
\item $E_{h(\bet)}$ is the Euler--Lagrange morphism associated to
     $h(\bet)$. More precisely, it can be proved \cite{Kol83,Kru73,Vit98}
     that any form $h(\bet)$ can be uniquely split into the sum
\bEq\label{first variation}
     (\pi^{2r+1}_{r+1})^*h(\bet)=E_{h(\bet)}-d_H(p_{h(\bet)})
\eEq
where $E_{h(\bet)} \in \Con^{1}_{(2r,0)}\wed\Var^{n}_{2r}$ and
$p_{h(\bet)} \in \Con^{1}_{(2r-1,r-1)} \wed \Var^{n-1}_{2r-1}$. Here
$E_{h(\bet)}$ and $d_H(p_{h(\bet)})$ are uniquely defined, but
$p_{h(\bet)}$, the \emph{momentum}, is not; see \cite{Kol83} for a
discussion.
\item $H_{h(\gam )}$ is the Helmholtz morphism
    \cite{Many99,KoVi02,Vit01} associated to the form $h(\gam)$. More
    precisely, it can be proved that any form
    $h(\gam)$ can be uniquely split into the sum
    \bEq\label{Helm}
    (\pi^{2r+1}_{r+1})^*h(\gam)=H_{h(\gam)}-d_H(q_{h(\gam)}),
    \eEq
    where $H_{h(\gam)} \in
    \Con^1_{(2r+1,r)}\wed\Con^{1}_{(2r+1,0)}\wed\Var^{n}_{2r}$ and
    $q_{h(\bet)} \in \Con^{2}_{2r-1} \wed \Var^{n-1}_{2r-1}$
    (here uniqueness is intended in the same way as above), with the
    additional condition that $H_{h(\gam)}$ is skew-adjoint in the first
    contact factor.

    We will use later on the special case: 

\noindent $[(\pi^{2r+1}_{r+1})^*\gam]$ $=$
     $[dE_{h(\beta)}]$ $\in$ $\cE_{n+1}(\For^{n+1}_{2r+1}/\Thd^{n+1}_{2r+1})$.  In this
    case, if we set $\eta\byd E_{h(\beta)}$, then $H_{d\eta}$ can be
    introduced as the skew-symmet\-rization of the morphism
    $\widetilde{H}_{d\eta}$ $\in$
    $\Con^1_{(4r+1,2r+1)}\otimes\Con^1_{(4r+1,0)}\otimes\Hor^n_{4r+1}$, which is
    characterized by
\bEq\label{eq:tildehelm}
    \cE(i_{j\Xi} d\eta) = j\Xi\con \widetilde{H}_{d\eta}
\eEq
(see \cite{KoVi02,Vit98} for details). We recall that $H$ and $\widetilde{H}$
    have the same kernel; more precisely,
$\Tilde{H}_{d\eta} = 0$ if and only if $H_{d\eta} = 0$ \cite{KoVi02}.
\end{enumerate}

Let us recall the coordinate expressions.
\begin{enumerate}
\item $h(\alpha)= A\,v_{\bX}$, where
     $A\in\mathcal{C}^{\infty}(J_{r+1}\bY)$ is a \lq special\rq{} 
polynomial in the derivatives of order
$r+1$ (in the sense of \cite{PaVi00}; see also \cite{Vit98})
    .
\item Being locally $h(\beta)=B_i^{\bsig}\,\omega^i_{\bsig}\wedge
     v_{\bX}$, we have the standard expression of the Euler--Lagrange
     morphism (see, \eg, \cite{Kol83,Kru73,Vin84})
\[
E_{h(\beta)}=(-1)^{\bsig}D_{\bsig}B^{\bsig}_i \, \omega^i\wedge v_{\bX}.
\]
\item In the simpler case
   $[(\pi^{2r+1}_{r+1})^*\gam]=
   [dE_{h(\beta)}]\in\cE_{n+1}(\For^{n+1}_{2r+1}/\Thd^{n+1}_{2r+1})$,
   that we will use later on, we have locally $dE_{h(\beta)}
   =\der_i^{\bsig}e_j\, \omega^i_{\bsig}\wedge \omega^j\wedge v_{\bX}$, where
   $e_j= (-1)^{\bsig}D_{\bsig}B^{\bsig}_j$, so that
\begin{gather}
\widetilde{H}_{dE_{h(\beta)}} = H_{i\,j}^{\bsig}\,
\omega^i_{\bsig}\otimes\omega^j\otimes v_{\bX},
\qquad
H_{dE_{h(\beta)}} = \frac{1}{2}H_{i\,j}^{\bsig}\,
\omega^i_{\bsig}\wedge\omega^j\wedge v_{\bX},
\label{eq:helm}
\\
H_{i\,j}^{\bsig} \byd\der_i^{\bsig}e_j
-\sum_{\abs{\brho}=0}^{2r+1-\abs{\bsig}}(-1)^{\abs{(\bsig,\brho)}}
\binom{\abs{(\bsig,\brho)}}{\abs{\brho}} D_{\brho}\der_j^{(\bsig,\brho)}e_i
\notag
\end{gather}
where $(\bsig,\brho)$ denotes the union of the multi--indices $\bsig$ and
$\brho$ (see \cite{KoVi02,Vit98,Vit01};
a local version has been also derived in \cite{Kru90,Sau89}).
\end{enumerate}

Next, we interpret the above spaces $\Var^{h}_r$, with $h\leq n+2$.
\begin{enumerate}
\item $\Var^k_r\byd\hL^k_r$, $k\leq n$. We recall that
$\hL^k_r=h(\Lambda^k_r)$.
     So, $\Var^n_r$ can be interpreted as
      the space of Lagrangians of order $r+1$ which are polynomials of
      \lq special type\rq{} \cite{PaVi00} in the higher order derivatives;
\item $\Var^{n+1}_r$ is the space of Euler--Lagrange morphisms
     associated to forms $h(\bet)$; it is a subspace of
     $\Con^1_0\wed\Hor^n_{2r+1}$ \cite{Vit98}.
\item $\Var^{n+2}_{r}$ is the space of Helmholtz (or Helmholtz--Sonin)
      morphisms associated to forms $h(\gam)$.
\end{enumerate}

We can read $\cE_k$ through the above isomorphisms $I_k$. We obtain
the exact sheaf sequence
\bEq\label{short}
\diagramstyle[size=1.8em]
\begin{diagram}
      0 & \rTo & \R & \rTo & \For^{0}_r & \rTo^{\cE_{0}} &
      \Var^1_r & \rTo^{\cE_{1}} & \dots & \rTo^{\cE_{n+1}} &
      \Var^{n+2}_r & \rTo & \dots
\end{diagram}
\eEq
It turns out \cite{Kru90,Vit98} that:
\begin{enumerate}
\item if $\mu\in\Var^k_r$, with $k\leq n-1$, and
      $\mu=h(\alp)=I_k(\alp)$, with $\alp\in\For^k_r$, then
      $\cE_k(\mu)=h(d\alp)$. Notice that
      $h({\pi^{r+1}_r}^*d\alp)=h((d_V+d_H)(h(\alp)+v(\alp)))=D_H(h(\alp))$,
      hence $\cE_k$ is equal to $d_H$ up to a pull-back;
\item if $\lam\in\Var^{n}_{r}$ then $\cE_{n}(\lam) \in\Var^{n+1}_{r}$
      coincides with the standard higher order Euler--Lagrange morphism
      associated with the Lagrangian $\lam$. We will simply write $\cE$ instead
      of $\cE_n$;
\item if $\eta\in\Var^{n+1}_{r}$ then $\cE_{n+1}(\eta)\in\Var^{n+2}_r$ is
      the Helmholtz morphism corresponding to the Euler--Lagrange morphism
      $\eta$. The exactness of \eqref{short} implies that $\cE_{n+1}(\eta)=0$ if
      and only if there exists (locally) a Lagrangian $\lam\in\Var^{n}_{r}$ such
      that $\cE(\lam)=\eta$, \ie, $\eta$ is \emph{locally variational}.
\end{enumerate}

\bDf
Let $k\leq n+1$. We say elements $\mu\in\Var^k_r$ to be \emph{variational
   forms}.
\eDf

\bRm\label{re:higher_var_forms}
We observe that the spaces $\For^{k}_r/\Thd^k_r$ with $k\geq n+2$ do not
have in the literature (to our knowledge) any interpretation in terms of
standard objects of the calculus of variations. In any case, there is a
representation $I_k$ also for such quotient spaces \cite{Vit01}. It comes from
the analogue representation for the variational sequence on infinite order jets
\cite[p. 192]{Many99}.  Furthermore, variational forms of degree $k>n+1$ will
not play any role in the rest of the paper.\END
\eRm

\bRm\label{re:nototdiv}
Due to $\cE_k\circ\cE_{k-1}=0$, Lagrangians of the form
$\lam\in\cE_{n-1}(\Var^{n-1}_r)$ are variationally trivial (\ie, they have
identically vanishing Euler--Lagrange equations).  Our aim in this paper is to
obtain an intrinsic model for the second and higher order variations. This is
achieved in the literature in several ways, each of which differs from the
others by a total divergence. Our viewpoint is different: we want to provide a
model which does not suffer the above arbitrariness.\END
\eRm

In view of the above Remark, we
factorize a part of the variational sequence as follows:
\bEq\label{factor}
\begin{diagram}
\Var^{n}_{r} && \rTo^{\cE} && \Var^{n+1}_{r}
\\
& \rdTo_{\tilde{\pi}} && \ruTo_{\bar{\cE}} &
\\
&& \bVar^{n}_{r} &&
\end{diagram}
\eEq
where $\bVar^{n}_{r}\byd\Var^{n}_{r}/\cE_{n-1}(\Var^{n-1}_{r})$, $\tilde{\pi}$
is the quotient map and $\bar{\cE}$ is the factor map.

\medskip

It is important to compute infinitesimal symmetries of objects in the
variational sequence. To do this, it is natural to look for vector
fields $X$ such that the \emph{standard Lie derivative operator $L_X$}
passes to the quotient into the variational sequence.

The jet prolongation $j\Xi$ of vertical vector fields $\Xi\colon
\bY\to V\bY$ preserves the contact structure on jets. Hence, it is
easy to see that the standard Lie derivative operator with respect to
$j\Xi$ preserves the contact sequence too. This yields the \emph{new operator}
$\cL_{j\Xi}$ on the elements of the variational sequence
\bEq\label{varLieder}
\cL_{j\Xi}\colon \Var^k_r\to\Var^k_r\colon \tau\mapsto
I_k([L_{j\Xi}\alpha]),
\eEq
where $[\alpha]=I^{-1}_k(\tau)$. The operator $\cL_{j\Xi}$ is
said to be the \emph{variational Lie derivative}~\cite{FPV98a}. Such
an operator allows us to recover several well--known formulae from the
calculus of variations (see, \eg, \cite{Kol81}) in a unique picture.
We have the following expressions~\cite{FPV98a}:
\begin{subequations}
\begin{enumerate}
\item if $0 \leq p\leq n-1$ and $\mu \in \Var^{p}_{r}$, then
\bEq\label{currLieder}
\cL_{j\xi}\mu = j\xi\rfloor d_{V}\mu;
\eEq
\item if $p = n$ and $\lam \in \Var^{n}_{r}$, then
\bEq\label{Lieder}
\cL_{j\xi}\lam = \xi\rfloor \cE(\lam)+
d_{H}(j\xi\rfloor p_{d_{V}\lam});
\eEq
\item if $p = n+1$ and $\eta \in \Var^{n+1}_{r}$, then
\bEq\label{helmholtz}
\cL_{j\xi} \eta = \cE(\xi\rfloor
\eta)+ j\xi\rfloor H_{d\eta}.
\eEq
\end{enumerate}
\end{subequations}
We remark that the operator $\cL_{j\Xi}\colon\Var^n_r\to\Var^n_r$
factorizes to $\bVar^n_r$, since the standard Lie derivative operator with
respect to $j\Xi$ commutes with $d_H$. This fact, combined with
\eqref{Lieder}, produces the further \emph{new quotient operator}
$\bar{\cL}_{j\Xi}$:
\bEq\label{quoLieder}
\bar{\cL}_{j\Xi}\colon
\bVar^n_r\to\bVar^n_r\colon \lambda \mapsto
\bar{\cL}_{j\Xi}\lambda\byd \Xi\con\cE(\lambda).
\eEq
It is clear that this operator can be interpreted as the
\emph{infinitesimal variation operator} of a Lagrangian \emph{up to total
   divergencies}. In the next section we provide a geometric model for such an
operator and its iterated applications.

\section{Variations of forms}
We shall here introduce the \emph{variation} of a form as \emph{infinitesimal
   multiparameter deformation}. This is realized by taking iterated Lie
derivatives of the form with respect to vertical vector fields. In this paper
we shall consider only vertical variations. In fact, as it is easy to realize
(see appendix 1), variations with respect to projectable vector fields do not
change the results, since they just add an horizontal differential (a total
divergence) which is not relevant for our results in view of Remark
\ref{re:nototdiv}.

\bigskip

Let $s :\bX\to\bY$ be a section and $i>0$.
Let $\Xi_{1},\ldots,\Xi_{i}$ be vertical vector fields on
$\bY$. Denote by $\psi^{l}_{t_{k}}$, with $1\leq l\leq i$, the
flows generated by $\Xi_{l}$. Then the map
\bEq
\Gam(t_{1},\ldots,t_{i})=
\psi^{i}_{t_{i}}\circ \ldots \circ \psi^{1}_{t_{1}} \circ s
\eEq
is said to be the \emph{$i$-th variation} of $s$ generated by
$(\Xi_{1},\ldots,\Xi_{i})$.

Let $\alp\in\For^{k}_r$ and let
$\Gam$ be an $i$--th variation of the section $s $. Then the map
\bEq \Del^{i}[\Gam;s ](\alp)
\byd \frac{\der^{i}}{\der t_{1}\ldots \der t_{i}}\big |_{t_{1},\ldots,
      t_{i}=0}( j_{r}\Gam(t_{1},\ldots,t_{i})^*\alpha)
\eEq
is said to be the \emph{$i$--th variation} of the form $\alp$ along the
section $s $.

The following Lemma states the relation between the $i$--th variation
of a form and its iterated Lie derivative.

\bLm\label{LIE}
Let $\alp:J_{r}\bY\to \owed{k}T^*J_{r}\bY$. Let
$\Gam$ be an $i$--th variation of the section $s $ generated by
variation fields $\Xi_{1}, \ldots, \Xi_{i}$.
Then we have
\bEq
\Del^{i}[\Gam;s ] (\alp) = (j_{r}s )^{*}L_{j_{r}\Xi_{1}}
\ldots L_{j_{r}\Xi_{i}}\alp.
\eEq
\eLm

\bPf
By the above definitions, we have
\begin{align*}
\Del^{i}[\Gam;s ]  (\alp)
&= \frac{\der^{i}}{\der t_{1}\ldots
\der t_{i}}\big |_{t_{1}\ldots
t_{i}=0}[(j_{r}(\psi^{i}_{t_{i}}\circ \ldots \circ \psi^{1}_{t_{1}} \circ
s ))^{*}\alp]
\\
&=(j_{r}s )^{*}\frac{\der^{i}}{\der t_{1}\ldots \der t_{i}}\big
|_{t_{1}\ldots
t_{i}=0}[(j_{r}\psi^{1}_{t_{1}})^{*}\circ
\ldots \circ (j_{r}\psi^{i}_{t_{i}})^{*}]\alp
\\
&=(j_{r}s )^{*}L_{j_{r}\Xi_{1}} \ldots L_{j_{r}\Xi_{i}} \alp ,
\end{align*}
where $\psi^{k}_{t_{k}}$ are the vertical flows
generated by $\Xi_{k}$ and we used the definition of prolongation of a
vertical vector field (see section \ref{sec:jets}).\QED
\ePf

  From the above considerations it follows that the definition of
variation can be given in terms of Lie derivatives with respect to
prolongations of vertical vector fields, without any reference to a
given section.

\bDf\label{def:variation2}
Let $\Xi_1$,\dots,$\Xi_i$ be vertical vector fields on $\bY$. Then the
\emph{variation} of a form $\alp\in\For^{k}_r$ is defined to be the operator
\beq
\Del^i[\Xi_1,\dots,\Xi_i](\alp)\byd
L_{j_r\Xi_1}(L_{j_r\Xi_2}(\dots(L_{j_r\Xi_i}\alp)\dots)).
\eeq
\eDf

\bRm\label{rem:trivial_variations}
The variation of an $(n+h)$-form along a section $s$ is clearly
zero. In fact, the above Lemma shows that the variation is obtained
through a pull-back on $\bX$ via $s$ and any $(n+h)$-form on
$\bX$ is zero. So, the definition of variation is trivial for $(n+h)$-forms.

Nonetheless, we can use the more general Definition 
\ref{def:variation2} of variation
  in all situations. But we loose the classical
interpretation of variation as `derivative along a parametrized family
of sections'.\END
\eRm
\section{Variations of Lagrangians}\label{sec:var_Lag}
In this section we restrict our attention to variations of elements of the
variational sequence, and, in particular, to Lagrangians. Our task is to
compute variations \lq up to variationally trivial forms\rq{}.  In other words,
we want to compute the \emph{quotient variations} of Lagrangians in the
variational sequence.  Indeed, this is straightforwardly permitted by our
definitions: $i$-th variations are made by Lie derivatives with respect to
prolonged vertical vector fields, and they pass to the quotient in the
variational sequence \cite{FPV98a}.

Moreover, we will devote special attention to variations along
sections which are \emph{critical} with respect to a given Lagrangian $\lam$,
\ie, sections $s$ such that $\cE(\lam)\circ js =0$.
\subsection{Quotient variation in the variational sequence}

Let $\alp\in\For^k_r$, with $k\leq n+1$, and $\Xi_1$,\dots,$\Xi_i$ be vertical
vector fields on $\bY$. We have
\begin{align*}
      I_k([\Del^{i}[\Xi_1,\dots,\Xi_i]\alp]) &= I_k([L_{j\Xi_{1}}(\dots
      (L_{j\Xi_{i}}\alp)\dots)])
      \\
      &=\cL_{j\Xi_{1}}(\ldots (\cL_{j\Xi_{i}}I_k([\alp]))\dots),
\end{align*}
where $\cL$ stands for the variational Lie derivative (see~\eqref{currLieder},
\eqref{Lieder}, \eqref{helmholtz}).

\bDf
The operator
\beq
\del^i[\Xi_1,\dots,\Xi_i]I_k([\alp])\byd
\cL_{j\Xi_{1}}(\ldots (\cL_{j\Xi_{i}}I_k([\alp]))\dots)
\eeq
is said to be the \emph{quotient variation} of the variational form
$I_k([\alp])\in\Var^{k}_r$ with respect to the vertical vector fields
$\Xi_1$,\dots,$\Xi_i$ on $\bY$.

If $s \colon\bX\to\bY$ is a (local) section of $\pi$, then
the form
\beq
(js)^*(\del^i[\Xi_1,\dots,\Xi_i]I_k([\alp]))
\eeq
is said the \emph{quotient variation} of $I_k([\alp])$ \emph{along $s$}.
\eDf

\bRm
The above definition of quotient variation can be applied to all quotient
spaces in the variational sequence (see Remark~\ref{re:higher_var_forms}).
But, if we want to calculate
variations of forms along sections, only the variations of forms
$[\alp]\in\Var^{k}_r$ with $k\leq n$ are non-trivial (see Remark
\ref{rem:trivial_variations}). In this work we will just devote ourselves
to variations of Lagrangians; nonetheless it would be interesting to
investigate variations of elements in $\Var^{k}_r$ with $k\leq n$.
\eRm

\subsection{Second variation}

In this subsection we fix a Lagrangian $\lam\in\Var^{n}_r$ and compute its
second quotient variation along a critical section $s$.

\medskip

Let $\Xi\colon\bY\to V\bY$ be a vertical vector field. It is natural
to introduce an improved quotient variation on the space $\bVar^n_r$ (see
\eqref{factor}). In fact, the operator $\bar{\cL}_{j\Xi}$
\eqref{quoLieder} is equal to the operator
$\cL_{j\Xi}$ \eqref{Lieder} up to \lq total divergencies\rq, \ie,
up to $\cE_{n-1}$-exact (variationally trivial) Lagrangians. We recall (see
Eq. ~\eqref{short}) that $\cE_{n-1}$ is equal to $d_H$ up to a pull-back,
or up to higher order variationally trivial terms.

\bDf
The operator
\begin{align*}
\bar{\del}^i[\Xi_1,\dots,\Xi_i][\lam] &\byd
\bar{\cL}_{j\Xi_{1}}(\ldots (\bar{\cL}_{j\Xi_{i}}\lam)\dots)
\\
&=\Xi_1\con\cE(\Xi_2\con\cE(\dots
\Xi_{i-1}\rfloor\cE(\Xi_i\con\cE(\lam))\dots))
\end{align*}
is said to be the \emph{quotient variation} of the Lagrangian $\lam$.

If $s \colon\bX\to\bY$ is a (local) section of $\pi$, then
the form
\beq
(js)^*(\bar{\del}^i[\Xi_1,\dots,\Xi_i][\lam])
\eeq
is called the \emph{quotient variation} of $[\lam]$ \emph{along $s$}.\END
\eDf

Of course, the \emph{first quotient variation} of $\lam$ is just
$\bar{\delta}^1[\Xi][\lam]=\Xi\con\cE(\lam)$.

Let $\Xi_1$, $\Xi_2$ be two vertical vector fields and let us consider the
\emph{second quotient variation} of $\lam$
\beq
\bar{\delta}^2[\Xi_1,\Xi_2][\lam]=\Xi_1\con\cE(\Xi_2\con\cE(\lam)).
\eeq

\bDf
We define the fibered morphism $\Xi_1\con\cE(\Xi_2\con\cE(\lam))$
 to be the \emph{Hessian morphism} associated with
     the Lagrangian $\lam$.
\eDf

We state our main result:

\bTh
The second quotient variation of a Lagrangian $\lam$ along a critical
section $s$ is equal to either one of the following intrinsic bundle
morphisms, which are self-adjoint:
\begin{enumerate}
\item the differential $V\cE(\lam)$ of $\cE(\lam)$ along the fibres of
  $\pi^{2r+1}$, also known as \emph{vertical differential}:
\[
V\cE(\lam)\colon J_{2r+1}\bY\to
  V^*J_{2r+1}\bY\otimes V^*\bY\otimes \wedge^n T^*\bX;
\]
\item the adjoint $V\cE(\lam)^*$ of the vertical differential:
\[
V\cE(\lam)^*\colon J_{2r+1}\bY\to
  V^*J_{4r+2}\bY\otimes V^*\bY\otimes \wedge^n T^*\bX
\]
(see \eqref{adjoint}).
\end{enumerate}
\eTh

\bPf
In coordinates we have $\Xi_1=\Xi_1^i\der_i$,
$\Xi_2=\Xi_2^j\der_j$ and $\lam = L\,\omega$. Then
\begin{align}
\bar{\delta}^2[\Xi_1,\Xi_2][\lam] &=
(-1)^{|\bsig|}\Xi_1^j\,D_{\bsig}(\der^{\bsig}_j
(\Xi_2^i\,\cE(L\ome)_i))\,\ome\notag
\\\label{secvar}
&=(-1)^{|\bsig|}\Xi_1^j\,D_{\bsig}(\der^{\bsig}_j\Xi_2^i\,\cE(L\ome)_i)\,\ome +
(-1)^{|\bsig|}\Xi_1^j\,D_{\bsig}(\Xi_2^i\,\der^{\bsig}_j\cE(L\ome)_i)\,\ome.
\end{align}
If $s$ is a critical section, then the first summand of
$\bar{\delta}^2[\Xi_1,\Xi_2][\lam]$ in the right-hand side of \eqref{secvar}
vanishes identically. The second summand admits an intrinsic interpretation. In
fact, we have $\cE^2(\lam)=0=\widetilde{H}_{\cE(\lam)}$ due to the property
$\cE_{n+1}\circ\cE_n=0$ of the variational sequence and the fact that $H$ and
$\widetilde{H}$ have the same kernel \cite{KoVi02}. Hence we have, from the
expression \eqref{eq:tildehelm}
\begin{align}
0&=j\Xi_1\con j\Xi_2\con \widetilde{H}_{\cE(\lam)}
\\
&=\Xi_1^i\,D_{\bsig}\Xi_2^j\,\der^{\bsig}_j\cE(\lam)_i\,\ome
- (-1)^{|\bsig|}\Xi_1^j\,D_{\bsig}(\Xi_2^i\,\der^{\bsig}_j\cE(L\ome)_i)\,\ome,
\label{secvar2}
\end{align}
so that the second summand of \eqref{secvar} is equal to the
first summand of \eqref{secvar2}, which is the vertical
differential $V\cE(\lam)$ of $\cE(\lam)$ (also known as
\emph{linearization} \cite{Many99}) contracted with the prolonged
fields $\Xi_1$, $\Xi_2$, namely
\bEq\label{jacolin}
\Xi_1\con j\Xi_2\con V\cE(\lam) =
\Xi_1^i\,D_{\bsig}\Xi_2^j\,\der^{\bsig}_j\cE(\lam)_i\,\ome\,.
\eEq
A comparison of the coordinate expressions shows that the second summand of
\eqref{secvar2} is equal to the adjoint of $V\cE(\lam)$ (see
\eqref{adjoint}). More precisely,
\[
\Xi_1\con j\Xi_2\con V\cE(\lam)=\Xi_1\con j\Xi_2\con V\cE(\lam)^*.\QED
\]
\ePf

\bCr
The morphism $V\cE(\lam)$ is symmetric along any critical section $s$, \ie
\beq
j_{2r+1}s^*(\Xi_1\con j\Xi_2\con V\cE(\lam))=j_{2r+1}s^*(\Xi_2\con
j\Xi_1\con V\cE(\lam)).
\eeq
\eCr

\bDf
We define the fibered morphisms:
$V\cE(\lam)^*=V\cE(\lam)$ to be the \emph{Jacobi morphism} associated
     with the Lagrangian $\lam$.
\eDf

It is not difficult to check that the above definition
recovers the definitions given by several authors up to `total
divergencies' (see, \eg, \cite{GH96,Olv93,Sau89}).

Our formulation however has the following advantages:
\begin{enumerate}
\item it is manifestly intrinsic (or covariant);
\item it holds for Lagrangians of arbitrary order, while in literature
     only the first order case is usually treated;
\item it allows an easy generalization to iterated variations of any order;
\item it allows an easy generalization to all spaces in the
variational sequence.
\end{enumerate}

\bRm
We can compare our approach with the one of Crampin
\cite{Cra00}. In that paper the Poincar\'e--Cartan form is used to achieve
an intrinsic formula for the second variation of a first-order
Lagrangian ($r=1$) in the case of one independent variable ($n=1$).
However, as is well-known, Poincar\'e--Cartan forms are no longer unique
in the case of many independent variables ($n>1$) and higher order
Lagrangians ($r>1$), so that our approach seems to be more suitable in the
general case.
\eRm

\section{Examples}

Here we show by simple but relevant examples that our definition of 
Jacobi morphism coincides with
the standard one and it can also be applied to higher order Lagrangians.

\bEx (Metric Lagrangian).
Here we shall derive the classical Jacobi equation for geodesics
within our framework.
Let $(\bQ, \bg)$ be a Riemannian manifold, with metric tensor $\bg
=g_{ab}d^{a}\ten d^{b}$. The Lagrangian
for geodesics is $\lam =\frac{1}{2}g_{ab}(q)\dot{q}^{a}\dot{q}^{b}dt$
and the Euler--Lagrange equation is given
by
\beq
\cE(\lam)_{a}=-[g_{ab}\ddot{q}^{b}+ \Gam_{abc}\dot{q}^{b}\dot{q}^{c}]=0\,,
\eeq
where $\Gam_{abc}$ are Christoffel symbols of the first kind.

The Jacobi equation is then obtained by evaluating the local
coordinate expression for the Jacobi morphism given
by \eqref{jacolin} for $\cE(\lam)_{a}$.
It is easy to see that the Jacobi morphism for geodesics is in fact
given in local coordinates by
\beq
V\cE(\lam)^{*}
= -
[\der_{a}g_{bc}\ddot{q}^{b}+\der_{a}\Gam_{bdc}\dot{q}^{b}\dot{q}^{d}]
\Xi^{a}_{2}\Xi^{c}_{1}+\\
+ [\der_{a}g_{bc}\dot{q}^{b}-
\der_{c}g_{ab}\dot{q}^{b}]\Xi^{a}_{2}\frac{d}{dt}\Xi^{c}_{1}+\\
- \der_{b}g_{ac}\dot{q}^{b}\Xi^{a}_{2}\frac{d}{dt}\Xi^{c}_{1} -
g_{ac}\Xi^{a}_{2}\frac{d^{2}}{d^{2}t}\Xi^{c}_{1}\,.
\eeq
Taking into account the Euler--Lagrange equation we get finally the
Jacobi equation
\beq
[\der_{a}g_{bc}\Gam^{b}_{de}
-\der_{c}\Gam_{dea}]\dot{q}^{d}\dot{q}^{e}\Xi^{c}_{1}-
2\Gam_{cba}\dot{q}^{b}\frac{d}{dt}\Xi^{c}_{1}-
g_{ac}\frac{d^{2}}{d^{2}t}\Xi^{c}_{1}=0\,,
\eeq
which can be recasted in the standard form
\beq
\nabla^{2}_{\dot{\gam}}\Xi_{1}+Riem(\Xi_{1},\dot{\gam},\dot{\gam}) =0\,,
\eeq
where $\gam$ is any geodesic curve, $\nabla^{2}_{\dot{\gam}}$ denotes
the second order covariant derivative along
the curve $\gam$ and $Riem(\Xi_{1},\dot{\gam},\dot{\gam})$ is the
Riemannian curvature tensor. This agrees of course with \cite{CaFr97b}.
\eEx

\bEx (Hilbert--Einstein Lagrangian).
Let $\dim\bX = 4$ and $\bX$ be orientable. Let
$Lor(\bX)$ be the bundle of Lorentzian metrics on $\bX$ (provided
that it has global sections). Local fibered coordinates on
$J_{2}(Lor(\bX))$ are
$(x^{\lam}; g_{\mu\nu}, g_{\mu\nu,\sig}, g_{\mu\nu,\sig\rho})$.

The Hilbert--Einstein Lagrangian is the form $\lam_{HE} \in
\Hor^{4}_{2}$ defined by $\lam_{HE} = L_{HE}\ome$, were
$L_{HE}=r\,\sqrt{\bg}$. Here $r: J_{2}(Lor(\bX)) \to \R$ is the function
such that, for any Lorentz metric $g$, we have $r\circ j_2g = R$,
being $R$ the scalar curvature associated with $g$ and $\bg$ the
determinant of $g$.

A direct computation of the Euler--Lagrange morphism shows
that $E_{d\lam_{HE}} = G$ $\byd$ $\textstyle{Ric}$ $-$ $\frac12\,R\,g \in
\Con^{1}_{(2,0)}$ $\wed$ $\Hor^{4}_2$, $\textstyle{Ric}$ being the Ricci
tensor of the metric $g$.

The Jacobi equations for the Hilbert--Einstein Lagrangian can be then
characterized as the
kernel of the adjoint of the linearization morphism $V\cE(\lam_{HE})$:
\begin{multline*}
V\cE(\lam_{HE})_{\alp}^{*\,\bet}=\frac{1}{2}[-\nabla_{\lam}\nabla^{\lam}
\Gam^{\bet}_{\alp}+r^{\bet\lam}\Gam_{\lam\alp}-
r_{\balp\lam}\Gam^{\lam\bet}-2R^{\bet}_{\rho\alp\lam}\Gam^{\rho\lam}+\\
+\del_{\alp}^{\bet}r_{\rho\lam}\Gam^{\rho\lam}+(G^{\bet}_{\alp}+
\frac{1}{2}s\del^{\bet}_{\alp})\Gam+
\nabla^{\bet}\nabla_{\lam}\Gam^{\lam}_{\alp}+\nabla_{\alp}(g^{\bet\gam}
\nabla_{\lam}\Gam^{\lam}_{\gam})-
\del^{\bet}_{\alp}\nabla_{\lam}(g^{\lam\gam}\nabla_{\rho}
\Gam^{\rho}_{\gam})]=0\,,
\end{multline*}
which coincide with the classical variation of the Einstein tensor (see, \eg,
\cite{Bla62}).

It is easy to realize that, along critical sections (\ie solutions of the Einstein equations), the Jacobi
morphism is in fact self-adjoint. This is also in accordance with 
\cite{Pom96}.
\eEx
\section{Conclusions}

We provided an intrinsic formalization of higher variations of a Lagrangian. In
the case of the second variation, we gave a new interpretation of the Hessian
and Jacobi morphism for Lagrangians of arbitrary order.

Some problems remain open at this point and will be investigated in the
future:
\begin{itemize}
\item It would be interesting to derive a formula for higher order variations,
   as well as higher order analogues of Hessian and Jacobi morphisms.
\item It would also be worth to compute variations of all variational forms
   (not only Lagrangians).
\item There are branches of Quantum Field Theory in which higher order
   variations play important roles, like \eg the Batalin--Vilkoviski theory
   \cite{HTQGS}. Such approaches still need a complete mathematical
   understanding. The above framework could be well-suited for that purpose: in
   \cite{HTQGS} the Batalin-Vilkoviski theory is formalized through 
jet bundles. Moreover, higher order
variations play in any case a role in the path integral approach to 
quantization.
\end{itemize}
\subsection*{Acknowledgments}

Thanks are due to I. Kol\'a\v r and A. M. Verbovetsky for many useful
discussions. The second author (M. P.) wishes to thank I. Kol\'a\v r for the
invitation and the kind hospitality at the Department of Algebra and Geometry
of the University of Brno in December 2000.

Commutative diagrams have been drawn by Paul Taylor's
\texttt{diagrams} macro package.
\section*{Appendix 1}

In this paper we considered only \emph{vertical} variations. More
general variations could be considered: in some problems of field theory (like
the computation of conserved currents) it is interesting to consider Lie
derivatives of Lagrangians with respect to \emph{projectable vector fields}. A
projectable vector field on $\bY$ is a pair $(\Xi,\bar{\Xi})$ such that
$\Xi$ is a vector field on $\bY$, $\bar{\Xi}$ is a vector field on
$\bX$ and $\Xi$ is a bundle morphism over $\bar{\Xi}$. In coordinates,
$\Xi=\bar{\Xi}^{\lambda}\der_\lambda+\Xi^i\der_i$, where
$\bar{\Xi}=\bar{\Xi}^{\lambda}\der_\lambda$.

However, such general variations do not modify the conclusions of our paper in
a significant way: their contribution to the variation of a 
Lagrangian is in fact a
total divergence, so that it has obviously to be neglected in our 
scheme. This fact is, of course,
well-known (see, \eg, \cite{Gri83}), but we recall it here for the sake
of completeness. The equation \eqref{Lieder} takes the more general form
  $\cL_{j_{r}\Xi}(\lam) = \Xi_{V}\rfloor \cE(\lam)+
d_{H}(j_{r}\Xi_{V}\rfloor p_{d_{V}\lam}+\Bar{\Xi}\rfloor \lam)$ (see, 
\eg, \cite{FPV98a}),
where $\Xi_V\byd\omega(\Xi)\colon J_1\bY\to V\bY$ is the \emph{vertical part}
of $\Xi$. In coordinates, $\Xi_V=(\Xi^i-y^i_\lambda\bar{\Xi}^\lambda)\der_i$.

It follows that our results of section \ref{sec:var_Lag} hold 
practically unchanged
  by just replacing vertical vector fields with vertical parts of projectable
vector fields.

\section*{Appendix 2}

Here we shortly recall the formulation of variational problems on jet spaces
\cite{GoSt73,Kol81,Kol83,Kru73,Kup80,Pal68,Sau89,Tul77} to help the reader to
connect the purely differential setting of variational sequences with
the classical integral presentation. Nonetheless, we stress the two
approaches (differential and integral) are completely independent,
even if the latter provided the motivation to the former from an
historical viewpoint.

Suppose that an $r$-th Lagrangian $\lambda\in\Hor^n_r$ is given. Then
the {\em action\/} of $\lambda$ on a section $s:\bU\to\bY$ ($\bU$ is
an oriented open subset of $\bX$ with compact closure and regular
boundary) is defined to be the real number
\beq
\int_{\bU}(j_rs)^*\lambda.
\eeq
A vertical vector field $\xi\colon
\bY\to V\bY$ defined on $\pi^{-1}(\bU)$ and vanishing on
$\pi^{-1}(\der\bU)$ is said to be a \emph{variation field}.
   A section $s\colon\bU\to\bY$ is said to be \emph{critical} if, for each
variation field with flow $\phi_t$, we have
$\delta\int_{\bU}(J_r\phi_t\circ j_rs)^*\lambda=0$, where $\delta$ is
the derivative with respect to the parameter $t$ and $J_r\phi_t\colon
J_r\bY\to J_r\bY$ is the jet prolongation of the flow $\phi_t$. It is
easy to see that the previous integral expression is equal to
$\int_{\bU}(j_rs)^*\text{L}_{u_r}\lambda=0$ for each variation field
$u$, where $u_r:J_r\bY\to VJ_r\bY$ is the $r$--th jet prolongation of
$u$ (see the first section).  Using equation \eqref{Lieder} together
with $L_{u_r}\lambda=i_{u_r}d\lambda$ and Stokes' Theorem, we find
that the above equation is equivalent to $\int_{\bU}(j_{2r}s)^*(i_u
E_{d\lambda})=0$ for each variation field $u$.  Finally, by virtue of the
fundamental Lemma of calculus of variations the above condition is
equivalent to $(j_{2r}s)^*E_{d\lambda}=0$, or, that is the same,
$E_{d\lambda}\circ j_{2r}s=0$.

Now the reason of the choice of the sheaf $\Thd^{k}_r$ (for $0\leq k\leq n$) as
the first non--trivial sheaf of the contact subsequence is clear: for $k=n$ the
sheaf $\Thd^{n}_r$ is made by forms which do not contribute to the action.  As
for the sheaf $\Thd^{n+1}_r$ it is easily seen that this is precisely the sheaf
of forms that give no contribution to the integral
$\int_{\bU}(j_{2r}s)^*i_uE_{d\lambda}$ when added to $E_{d\lambda}$.

\end{document}